\documentclass[12pt,preprint]{aastex}
\usepackage{natbib}

\shorttitle{Grain growth and mass decrease in disks in MBM~12}

\shortauthors{Hogerheijde et al.}

\begin{document}

\title {Indications for grain growth and mass decrease in cold dust disks
around Classical T~Tauri stars in the MBM~12 young association}

\author{Michiel R. Hogerheijde\altaffilmark{1},
	Doug Johnstone\altaffilmark{2},
	Isamu Matsuyama\altaffilmark{3},
	Ray Jayawardhana\altaffilmark{4},
    and James Muzerolle\altaffilmark{1}}

\altaffiltext{1}{Steward Observatory, The University of Arizona, 933
   N. Cherry Ave, Tucson, AZ 85721, U.S.A.; mhogerheijde@as.arizona.edu,
   jamesm@as.arizona.edu}
\altaffiltext{2}{Herzberg Institute of Astrophysics, National
   Research Council of Canada, 5071 West Saanich Road, Victoria, BC
   V9E 2E7, Canada; doug.johnstone@nrc-cnrc.gc.ca}
\altaffiltext{3}{Department of Astronomy and Astrophysics, University
   of Toronto, Toronto, ON M5S 3H8, Canada; isamu@astro.utoronto.ca}
\altaffiltext{4}{Department of Astronomy, University of
   Michigan, 830 Dennison Building, Ann Arbor, MI 48109, U.S.A.;
   rayjay@umich.edu}

\begin{abstract}
We report detection of continuum emission at $\lambda=850$ and
450~$\mu$m from disks around four Classical T~Tauri stars in the
MBM~12 (L1457) young association. Using a simple model we infer masses
of 0.0014--0.012~M$_\odot$ for the disk of LkH$\alpha$~263~ABC,
0.005--0.021~M$_\odot$ for S18~ABab, 0.03--0.18~M$_\odot$ for
LkH$\alpha$~264~A, and 0.023--0.23~M$_\odot$ for LkH$\alpha$~262. The
disk mass found for LkH$\alpha$~263~ABC is consistent with the
0.0018~M$_\odot$ inferred from the scattered light image of the
edge-on disk around component C. Comparison to earlier $^{13}$CO line
observations indicates CO depletion by up to a factor 300 with respect
to dark-cloud values. The spectral energy distributions (SED) suggest
grain growth, possibly to sizes of a few hundred $\mu$m, but our
spatially unresolved data cannot rule out opacity as an explanation
for the SED shape. Our observations show that these T~Tauri stars are
still surrounded by significant reservoirs of cold material at an age
of 1--5~Myr. We conclude that the observed differences in disk mass
are likely explained by binary separation affecting the initial
value. With available accretion rate estimates we find that our data
are consistent with theoretical expectations for viscously evolving
disks having decreased their masses by $\sim 30$\%.
\end{abstract}


\section{Introduction\label{s:intro}}

The majority of low-mass stars emerge from their parental clouds
surrounded by disks of 0.001--0.3~M$_\odot$
\citep{beckwith:nature}. At ages of a few Myr, these disks appear to
evolve rapidly from optically thick at near- and mid-infrared and
detectable at (sub)millimeter wavelengths, to undetectable at all
wavelengths \citep{robberto:sf99, rayjay:mirtwa, carpenter:ic348,
haisch:freq_clusters, luhman:mbm12, hartmann:agespread,
duvert:exdisk}. The physics behind this transition and its timescale
holds clues about the planet formation process
\citep{clarke:disk_uv,matsuyama:viscousdisk, matsuyama:halt,
armitage:dispersion}. The previous references focus on inner disk
material traced by infrared excess. Much less is known about colder
material further from the star, even though this encompasses the bulk
of the mass. This Letter investigates the presence of cold material
around several members of the MBM~12 young association.

Judging from the relative occurrences of K- ($\sim 20$\%) and L- and
N-band ($71\pm 32$\%) infrared excess, the MBM~12 (L1457) young
association is suspected to be at the very stage where disks start to
disappear \citep{luhman:mbm12,rayjay:mbm12_mir}. After several
unsuccessful attempts (\citealt{pound:mbm12}, \citealt{mrh:mbm12}),
\citet{itoh:mbm12} recently reported detection at 1 and 2~mm of
continuum emission around two or possibly three Classical T~Tauri
stars in this 1--5~Myr old association \citep{luhman:mbm12},
indicating the presence of $\sim 0.05$~M$_\odot$ of cold material
around each object. This Letter increases the number of detections of
cold dust to four objects (\S \ref{s:results}), including the recently
identified edge-on disk source LkH$\alpha$~263~C
\citep{rayjay:mbm12_ao}. By extending the wavelength coverage into the
submillimeter we can fit the spectral energy distributions (SEDs), and
gain more robust disk-mass estimates and insight into grain growth (\S
\ref{s:models}). The Letter concludes with a discussion of the
inferred mass range in terms of multiplicity
\citep{rayjay:mbm12_ao,chauvin:mbm12,brandeker:keckao} and
disk-dispersal models (\S \ref{s:naturenurture}).

\section{Observations\label{s:obs}}

The observations were obtained with the {\it Submillimeter Common User
Bolometer Array\/} (SCUBA) \citep{holland:scuba} on the James Clerk
Maxwell Telescope\footnote{The JCMT is operated by the Joint Astronomy
Centre in Hilo, Hawaii, on behalf of the parent organizations PPARC in
the United Kingdom, the National Research Council of Canada and The
Netherlands Organization for Scientific Research.} (JCMT) on 2002
December 12 under excellent weather conditions. Typical opacities at
225~GHz were 0.04--0.06. We obtained photometry of four systems:
LkH$\alpha$~262, the triple LkH$\alpha$~263 (ABC), LkH$\alpha$~264
(A), and the triple S18 (ABab); Table~\ref{t:obs} lists coordinates
and observing details. The employed two-bolometer photometry technique
allowed for increased observing efficiency, with a chop throw of
$52{\farcs}4$ in Naysmith coordinates. Individual integrations of 30~s
were repeated for the totals listed in the table. To confirm the
unresolved nature of the emission we obtained a 64-point jiggle map
centered on LkH$\alpha$~262 also containing LkH$\alpha$~263 (ABC).
While providing spatial information, the jiggle map is less sensitive
than the single-point photometry data.

The standards HL~Tau, CRL~618, and Uranus provided focus checks and
flux calibrations; the nearby quasar 0235+164 served as pointing
source every $\sim 2$ hours.  Pointing accuracy was good with
excursions of less than a few arcsec. However, uncorrected pointing
errors can still affect the photometry in the $8''$ beam at
450~$\mu$m. In spite of a $3''$ dither included in the photometry,
450~$\mu$m fluxes of LkH$\alpha$~262 from photometry are lower by 60\%
than those from the jiggle map (\S \ref{s:results}). We conclude that
uncorrected pointing offsets and calibration uncertainties at short
wavelengths due to the imperfect beam shape of the JCMT resulted in
450~$\mu$m photometry results that are strict lower limits to the
actual source flux. We include a +60\% error in the uncertainty of the
reported 450~$\mu$m results. The 850~$\mu$m photometry and the jiggle
maps are unaffected.

\section{Results\label{s:results}}

All four objects, containing a total of eight (known) stars, show
emission at 850 and 450~$\mu$m (Table~\ref{t:flux}). The jiggle map
(Fig.~\ref{f:map}) indicates that the emission is unresolved and
confined to the source position, and is not structure in the
cloud. Because of the higher noise level, only LkH$\alpha$~262 is
detected in Fig.~\ref{f:map} while LkH$\alpha$~263 (ABC) remain
undetected. The separation of $17''$ between LkH$\alpha$~262 and 263
(ABC) is large enough that $<1$--2~mJy spill over at both wavelengths
is expected, based on archival beam profiles.

High signal-to-noise photometry and jiggle-map data of LkH$\alpha$~262
are consistent at 850~$\mu$m but discrepant at 450~$\mu$m with
respective fluxes of 166.2~mJy and 263.1~mJy. Uncorrected pointing and
calibration errors at 450~$\mu$m are likely to blame (\S\ref{s:obs});
the jiggle-map flux is extracted after Gaussian profile fitting and
does not suffer from pointing offsets. The reconcile the measurements,
we include a +60\% error bar in Table~\ref{t:flux} and stress that the
450~$\mu$m photometry values are strict lower limits. The
850--450~$\mu$m spectral indices $\alpha$ consequently contain a large
uncertainty (Table \ref{t:flux}). At their high end, corresponding to
the high end of allowed 450~$\mu$m fluxes, the indices
($\alpha=$1.5--2.5) are consistent with emission from cool ($<30$~K)
and coagulated dust grains.  The spectral index $\alpha=\beta+\gamma$,
where the index of the dust emissivity $\beta$ decreases when grains
coagulate (e.g., \citealt{ossenkopf:kappa, pollack:kappa}) and where
the slope of the Planck function $\gamma$ falls below the value of 2.0
outside the Rayleigh-Jeans limit when $T\lesssim h\nu/k = 32$~K at
450~$\mu$m.

\section{Disk models and masses\label{s:models}}

Fig.~\ref{f:seds} plots our data and values from \citet{pound:mbm12},
\citet{mrh:mbm12}, and \citet{itoh:mbm12}; the +60\% error bars to the
450~$\mu$m photometry are explicitly included. From these SEDs we can
infer the mass of cold material. The 850~$\mu$m fluxes trace the
absolute amount of material, while the SEDs help to constrain
important model parameters. We choose the flared-disk model of
\citet{chiang:disksed} to describe the disks. We use the disk mass as
our only free parameter. The disk temperature distribution is fixed by
scaling to the stellar luminosity $\propto (L_\star/L_0)^{-0.25}$,
where $L_0=1.42$~L$_\odot$ \citep{chiang:disksed}. Source luminosities
$L_\star$ are from \citet{luhman:mbm12}. We neglect any changes in
disk structure due to changes in mass and temperature. Following
recent reassessments of the distance to MBM~12
\citep{luhman:mbm12,andersson:l1457} we adopt $275\pm 60$~pc. We
assume an average inclination of $60^\circ$.  Because most of the flux
comes from optically thin regions, only near edge-on orientations
would change our results significantly. 

The final model parameter is the dust emissivity $\kappa_\nu$: its
absolute value and its variation with wavelength. Often, $\kappa_\nu$
is parameterized as $\kappa_\nu=\kappa_0 (\nu_0)^\beta$ with
$\kappa_0=10.$~cm$^{2}$~g$^{-1}$ (dust) and $\nu_0=10^{12}$~Hz
\citep{hildebrand:kappa}. From disk SED fitting, low values of $\beta$
are commonly found, 0--1 \citep{beckwith:beta} and interpreted as
evidence for grain growth; \citet{beckwith:ppiv} point out that grain
mineralogy can affect $\beta$ and explain the low values. More
accurate descriptions of $\kappa_\nu(\nu)$ follow from calculations of
grain growth in astrophysical environments employing realistic
mineralogies (e.g., \citealt{ossenkopf:kappa, pollack:kappa,
beckwith:ppiv}). These models circumvent having to choose a value for
$\kappa_0$, but still show show significant variation. We adopt a dual
approach: we fit the SEDs with a parameterized $\kappa_\nu$ to
constrain $\beta$ and learn about possible grain growth, and then fit
the SEDs with several physical grain-growth models to derive accurate
disk masses with realistic uncertainties.

With $\kappa_\nu = \kappa_0 (\nu_0)^\beta$, $\chi^2$ minimization of
the SED fit yields $\beta = 0.5 \pm 0.3$ for LkH$\alpha$~262 and $0.8
\pm 0.5$ for LkH$\alpha$~264; the error bars for LkH$\alpha$~263 and
S18 are too large to constrain $\beta$. For this fit we have used all
detections and upper limits at wavelengths $\ge 450$~$\mu$m. At the
wavelengths of the infrared N-band and shorter, the flux is dominated
by very small amounts of hot material and may contain contributions
from silicate emission; both depend strongly on the model
parameters. This does not affect our results: $<1$\% of the 450~$\mu$m
flux originates from the $<4$~AU region that contributes $>96$\% of
the 10~$\mu$m emission. The inferred values $\beta = 0.5 \pm 0.3$ and
$0.8 \pm 0.5$ are similar to those commonly found in T~Tauri disks,
and we interpret them as indications for grain growth in the MBM~12
disks.

Few of the grain-growth models of \citet{ossenkopf:kappa} and
\citet{pollack:kappa} have $\beta<1$. Of the models by
\citeauthor{ossenkopf:kappa}, only strongly coagulated grains without
ice mantles show such low $\beta$. The authors note that such grains
are unlikely in disk environments and we follow their advice and avoid
this particular class. Of the models by \citeauthor{pollack:kappa},
$\beta<1$ is only found for average grain sizes $\gtrsim
1$~cm. Table~\ref{t:flux} lists the range of disk masses found by
$\chi^2$ minimization. These vary from 0.0014-0.012~M$_\odot$ for
LkH$\alpha$~263 to 0.023--0.23~M$_\odot$ for LkH$\alpha$~262. The
allowed range for each source reflects the uncertainty in the dust
opacity models. These these ranges overlap, but differences in disk
mass between sources are robust if we assume similar dust properties
in all objects.

Our modeling makes no assumptions about the opacity. We find that
$<15$\% of the 450~$\mu$m flux and $<7$\% at 1.3~mm originates from
opaque regions ($\tau\ge 3)$. Only much smaller disks ($\lesssim
30$~AU) have $\sim 85$\% of the 450~$\mu$m flux coming from optically
thick regions (45\% at 1.3~mm). For such disks the SED slope can be
explained by opacity instead of grain growth, and submillimeter fluxes
no longer trace mass. Only spatially resolved data can settle this
issue.

\citet{rayjay:mbm12_ao} infer 0.0018~M$_\odot$ for LkH$\alpha$~263~C
from modeling the scattered light, depending on the assumed dust
properties. Our values of 0.0014--0.012~M$_\odot$ are consistent with
their findings. An unknown fraction of our inferred mass may reside in
disks around the companions A and B. Spatially resolved data and
detailed modeling are required to further characterize the disk of
LkH$\alpha$~263~C.

$^{13}$CO 2--1 measurements \citep{mrh:mbm12} indicate that CO is
depleted with respect to the dark cloud value of CO/H$_2$=$2\times
10^{-4}$ by factors up to 300. Large depletion levels are found in
disks around T~Tauri stars (e.g.,
\citealt{dutrey:ggtau,dutrey:tts13co}) and expected theoretically
(e.g., \citealt{aikawa:chem2d,willacy:photodisk}).

\section{Nature or nurture\label{s:naturenurture}}

Although our sample is small, with only eight stars in four unresolved
systems, interesting conclusions can be reached about the disk
evolution. Our detections show that at least the Classical T~Tauri
stars in MBM~12 (1--5~Myr) have disks in the same mass range as the
younger ($\sim 1$~Myr) Taurus and $\rho$~Ophiuchus regions
(0.001--0.3~M$_\odot$; \citealt{beckwith:nature}). While
\citet{carpenter:ic348} shows that 0/95 stars in IC~348 ($\sim 2$~Myr)
have disk masses $>0.025$~M$_\odot$, two out of four of our MBM~12
systems do (or $>17$\% of all twelve MBM~12 systems); in Taurus, a
comparable fraction of 14\% of stars have disks
$>0.025$~M$_\odot$. \citet{luhman:mbm12} argues that the K- and L-band
excess suggest significant disk dispersal in MBM~12. Our sample does
not directly address this issue because 3/4 objects have K-band excess
(S18 does not).

\citet{jensen:binaries} showed that binaries with separations
$<50$--100~AU have reduced disk mass or no disks, compared to wider
binaries and single stars. Our data follow that trend, taking into
account that the relative disk masses are robust (\S\ref{s:models}):
LkH$\alpha$~262: $15{\farcs}3=4208$~AU and 0.023--0.23~M$_\odot$;
LkH$\alpha$~264~A: $9{\farcs}2=2530$~AU and 0.03--0.18~M$_\odot$;
S18~ABab: $0{\farcs}75=206$~AU (A--B), ${0\farcs}063=17$~AU (Ba--Bb)
and 0.005--0.025~M$_\odot$; and LkH$\alpha$~263~ABC:
$4{\farcs}1=1128$~AU (AB--C), $0{\farcs}41=113$~AU (A--B) and
0.0014--0.012~M$_\odot$ (binary separations from
\citealt{rayjay:mbm12_ao}, \citealt{chauvin:mbm12} and
\citealt{brandeker:keckao}). This suggests that the presence of a
close binary (environment) determines the total amount of disk
material.

If environment determines (initial) disk mass, can we still infer
anything about disk dispersal from the amount of material? In a
viscouly evolving disk, disk mass and the accretion rate onto the
central star are linked \citep{clarke:disk_uv,
matsuyama:viscousdisk}. Fig.~\ref{f:evol} plots `evolutionary tracks'
of disks with initial masses of 0.001, 0.01, and 0.1~M$_\odot$ around
a 0.5~M$_\odot$ star, following \citet{matsuyama:viscousdisk}. This
model assumes a standard $\alpha$-disk description with
$\alpha=10^{-3}$, a central star with $M_\star = 0.5$~M$_\odot$ and
$R_\star = 2$~R$_\odot$, and a constant ionizing flux from the star
with time-dependent contribution from accretion. The disk removal rate
only depends weakly on the stellar parameters, and any uncertainties
are insignificant compared to our observational error bars.

Muzerolle et al. (in prep.)  report estimates of mass accretion rates
onto three of our objects from Br$\gamma$ line measurements, using the
calibration of \citet{muzerolle:brgamma}. They find upper limits of
(4--5)$\times 10^{-9}$~M$_\odot$~yr$^{-1}$ for LkH$\alpha$~263 and
264, and a rate of $(2\pm 1)\times 10^{-9}$~M$_\odot$~yr$^{-1}$ for
S18. These rates are comparable to those inferred for IC~348
\citep{liu:ic348} and at the low end of the range found in Taurus,
which has a median of $10^{-8}$~M$_\odot$~yr$^{-1}$. The errors
include uncertainties in the observations and the stellar mass,
assumed to be 0.5~M$_\odot$, but not the distance. Because both
accretion rate and disk mass depend on the distance squared, its
uncertainty does not enter into the values of Fig.~\ref{f:evol}; $\dot
M_{\rm acc}$ also depends on $M_\star/R_\star$, which we assume
constant.

The disk mass and accretion rate of S18 are consistent with the
dispersal model for the age range of MBM~12 (1--5~Myr; shaded in
Fig.~\ref{f:evol}), suggesting that the object has lost one-third of
its initial disk mass of $\sim 0.016$~M$_\odot$. The upper limit on
the accretion rate of LkH$\alpha$~264 also agrees with the model at
1--5~Myr. The agreement improves if the disk mass is decreased
somewhat, which is expected when dust grains have indeed grown to a
few hundred $\mu$m as suggested by the SED. Grain sizes comparable to
observing wavelength are efficient emitters, reducing the required
mass to fit the observed flux. The limits on LkH$\alpha$~263 do not
provide useful constraints.

In summary, we conclude that the Classical T~Tauri stars in MBM~12
still have significant reservoirs of cold dust in circumstellar
disk. There are indications for grain growth up to several hundred
$\mu$m in these disks, but spatially resolved observations are
required to rule out opacity as an explanation for the flat spectral
slopes. And while differences in disk mass are likely dominated by
environment (binary separation; `nature'), available accretion rates
and detected disk masses are consistent with a disk dispersal scenario
(`nurture') with one-third of the mass already lost.

\acknowledgments We thank our TSSs J.~Kemp and J.~Hoge for excellent
support during our observations. The staff of the JCMT and the JAC are
thanked for their hospitality. RJ acknowledges support in part through
NASA Origins Grant NAG5-11905. I.M. acknowledges support from the
National Reserach Council, Canada. The referee is thanked for a
careful reading of our manuscript and insightful comments.

\newpage

\figcaption[]{Jiggle maps at (a) 850~$\mu$m and (b) 450~$\mu$m
containing four (known) young stars indicated by symbols:
LkH$\alpha$~263~C, LkH$\alpha$~263~A+B, and LkH$\alpha$~262 (left to
right). The emission around LkH$\alpha$~262 is unresolved. Emission
from LkH$\alpha$~263~ABC is undetected at the noise level of
7~mJy~beam$^{-1}$ (850~$\mu$m) and 22~mJy~beam$^{-1}$
(450~$\mu$m).\label{f:map}}

\figcaption[]{Spectral energy distributions (SEDs) of the four sources
detected with SCUBA. Filled symbols are SCUBA data, open symbols data
from \citet{itoh:mbm12}, star symbols show N-band excess
\citep{rayjay:mbm12_ao}, and $2\sigma$ upper limits are from
\citet{pound:mbm12}, \citet{mrh:mbm12}, and \citet{itoh:mbm12}. The
broad shaded curves show the model SEDs and their uncertainty due to
variations in adopted dust emissivities. Emission at $\lesssim
20$~$\mu$m originates from small amount of warm dust and is highly
model dependent.\label{f:seds}}

\figcaption[]{`Evolutionary tracks' of viscous accretion disks from
\citet{matsuyama:viscousdisk} (solid curves), for a stellar mass of
0.5 M$_\odot$ and initial disk masses of 0.1, 0.01, and 0.001
M$_\odot$. Dashed lines are isochrones at $1\times 10^4$, $4\times
10^4$, $1.6\times 10^5$, ...~yr. At these ages, the disks are reduced
to 99\%, 97\%, 91\%, 75\%, 50\%, and 27\% of their initial mass.  The
shaded area is the inferred age range of 1--5~Myr of MBM~12. Upper
limits are $2\sigma$ values.\label{f:evol}}

% Tables:

\newpage

\begin{deluxetable}{lrrrr}
\tablecaption{Observations\label{t:obs}}
\tablecolumns{5}
\tablehead{
 & \colhead{$\alpha$(2000)} & \colhead{$\delta$(2000)} & 
   \colhead{Observing} & \colhead{Integration} \\
\colhead{Source} & \colhead{(h m s)} & \colhead{($^\circ$ $'$ $''$)} & 
   \colhead{Mode} & \colhead{Time (s)}
}
\startdata
LkH$\alpha$262     & 02 56 07.9 & 20 03 25 & photom & 1200 \\
                   &            &          & jiggle & 4800 \\
LkH$\alpha$263 ABC\tablenotemark{a} 	     
                   & 02 56 08.7 & 20 03 41 & photom & 4800 \\
LkH$\alpha$264 A   & 02 56 37.5 & 20 05 38 & photom & 1200 \\
S18 ABab           & 03 02 21.1 & 17 10 35 & photom & 1200 \\
\enddata
\tablenotetext{a}{Centered on LkH$\alpha$263~C, $4{\farcs}1$ from AB.}
\end{deluxetable}

\clearpage

\begin{deluxetable}{lrrrr}
\tablecaption{Results\label{t:flux}}
\tablecolumns{5}
\tablehead{
 & \colhead{F(850 $\mu$m)} & \colhead{F(450 $\mu$m)} & 
   \colhead{Spectral} & \colhead{Disk Mass} \\
\colhead{Source} & \colhead{(mJy)} & \colhead{(mJy)} & 
   \colhead{Index $\alpha$} & \colhead{($10^{-2}$ M$_\odot$)}
}
\startdata
LkH$\alpha$~262 (photom) &  $97.2 \pm 3.4$ & $166.2 ^{+99.7}_{-11.8}$  & 
   0.67--1.64 & 0.023--0.23\tablenotemark{a}  \\
LkH$\alpha$~262 (jiggle) &  $95.4 \pm 7$   & $263.1 \pm 22$            & 
   1.35--1.84 &              \\
LkH$\alpha$~263ABC       &   $7.5 \pm 1.8$ &  $26.7 ^{+16.0}_{-7.2}$   & 
   1.16--3.17 & 0.0014--0.012\tablenotemark{a} \\
LkH$\alpha$~264A         & $136.7 \pm 3.4$ & $214.1 ^{+128.5}_{-11.9}$ & 
   0.58--1.48 & 0.03--0.18\tablenotemark{a}    \\
S18 ABab                 &  $21.6 \pm 3.3$ &  $65.2 ^{+39.2}_{-11.6}$  & 
   1.20--2.74 & 0.005--0.025\tablenotemark{a}  \\
\enddata
\tablenotetext{a}{The lower mass limits corresponds to dust opacities
from \citet{ossenkopf:kappa} with ice mantles and $>10^6$~yr of grain
growth; the upper mass limits to opacities from \citet{pollack:kappa}
with grain sizes $\sim 1$~cm.}
\end{deluxetable}

% Figures:

\newpage

\begin{figure}
\figurenum{\ref{f:map}}
\epsscale{0.7}
\plotone{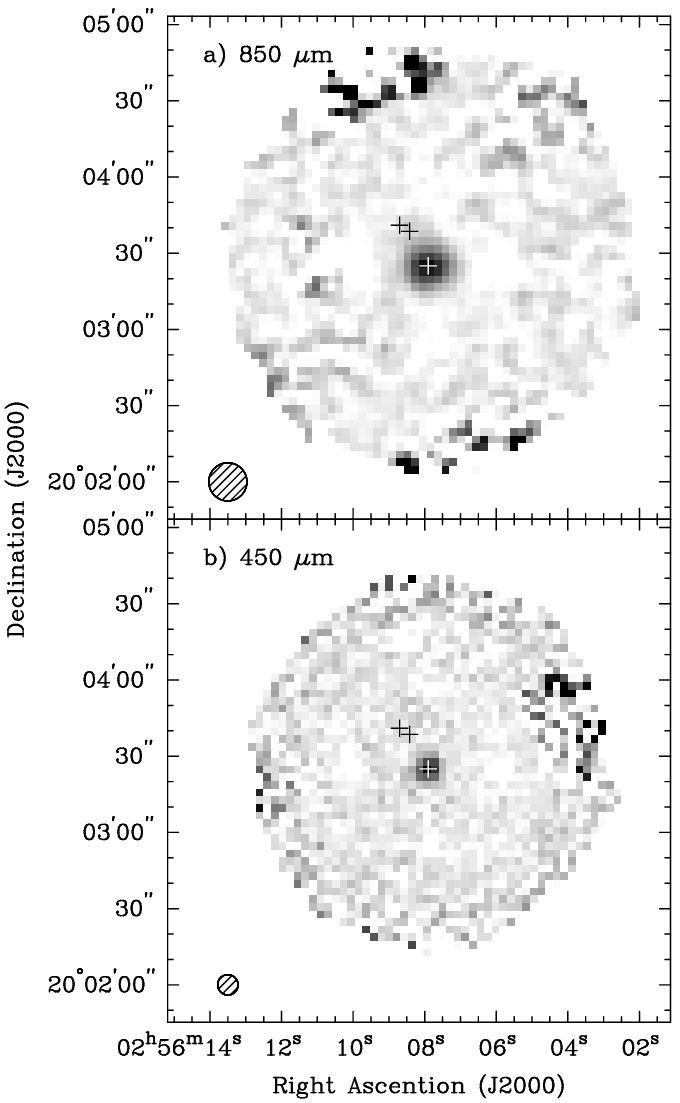}
\caption{}
\end{figure}

\begin{figure}
\figurenum{\ref{f:seds}}
\epsscale{0.7}
\plotone{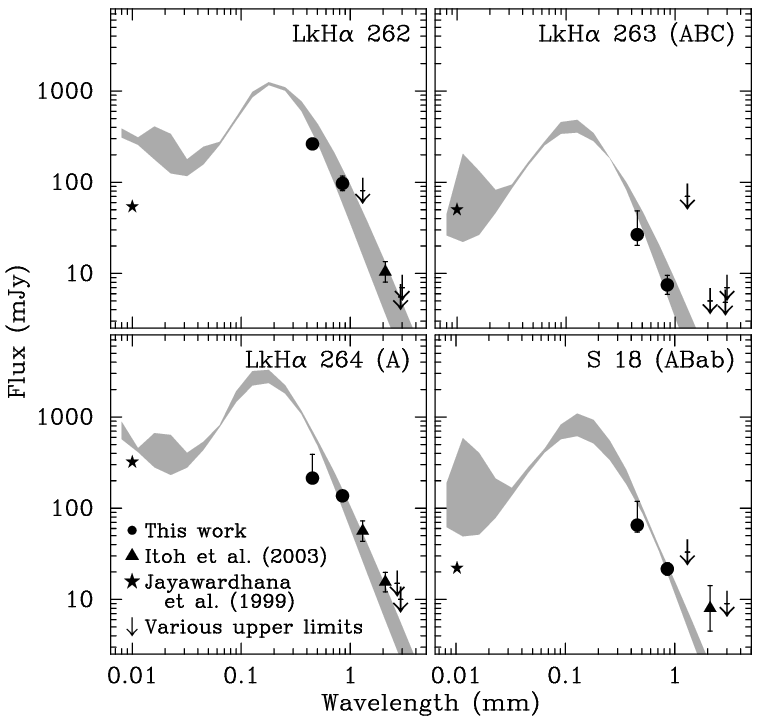}
\caption{}
\end{figure}

\begin{figure}
\figurenum{\ref{f:evol}}
\epsscale{0.7}
\plotone{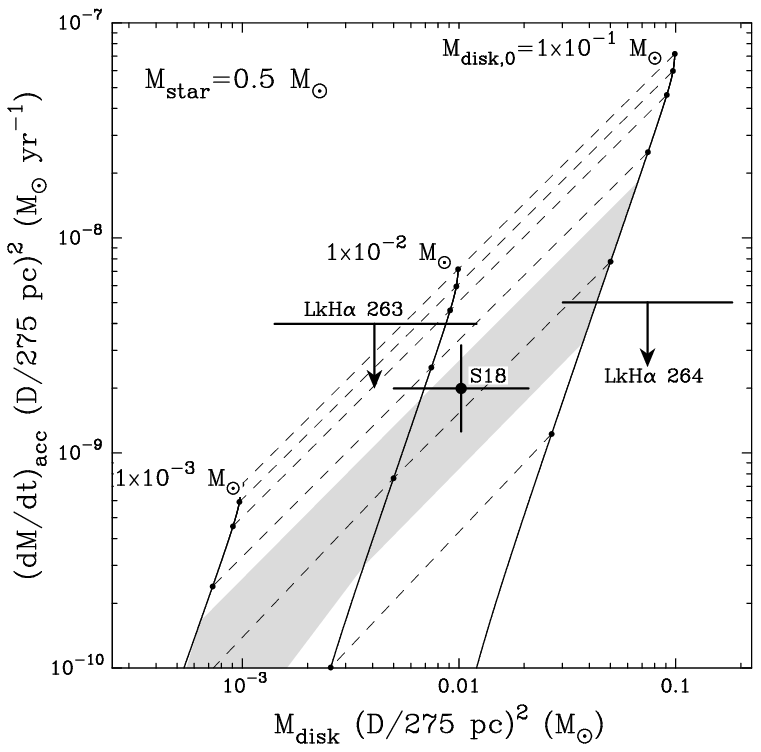}
\caption{}
\end{figure}

\end{document}